\documentstyle[preprint,aps]{revtex}
\begin{document}
\topmargin=-1cm
\vsize=27cm

\draft

\title{ 
{\Large UNIVERSIT\'E DE FRIBOURG SUISSE}\\ 
{\small INSTITUT DE PHYSIQUE TH\'EORIQUE}\\ 
\includegraphics{siglefr.eps}
\vspace*{2.0cm}
Interaction-Induced Enhancement and Oscillations of the 
Persistent Current}
\author{C.~A.~Stafford$^{1}$
and D.\ F.\ Wang$^2$}
\address{$\mbox{}^1$Institut de Physique 
Th\'{e}orique, Universit\'{e} de Fribourg,
CH-1700 Fribourg, Switzerland}
\address{$\mbox{}^2$Institut de Physique Th\'eorique, EPF 
Lausanne, CH-1015 Lausanne, Switzerland}
\maketitle

\begin{center}
\fbox{UF-IPT 1996 / 11-001}
\end{center}

\begin{abstract}
The persistent current $I$ in integrable models of multichannel
rings with both short- and long-ranged interactions is investigated.
$I$ is found to oscillate in sign and increase in
magnitude with increasing interaction strength 
due to interaction-induced correlations in the currents
contributed by different channels. 
For sufficiently strong interactions, the contributions of all channels
are found to add constructively, leading to a giant enhancement of $I$.
Numerical results confirm that this parity-locking effect is robust 
with respect to subband mixing due to disorder.
\end{abstract}

\tighten
\widetext

The absence of macroscopic persistent currents in normal metal rings even
at $T=0$ is a consequence of a parity effect stemming from Fermi statistics,
first discussed by Byers and Yang \cite{byang}:  For $N$ spinless electrons
in a purely one-dimensional (1D) ring, 
the persistent current $I$ is diamagnetic
if $N$ is even and paramagnetic if $N$ is odd, 
independent of disorder and interactions 
\cite{leggett,loss}, and takes a maximum value $I_0 =e v_F/L$
for a clean ring, where $v_F$ is the Fermi velocity and $L$ the 
circumference of the ring.  In a ring with many independent channels, 
$I$ is the
sum of many such diamagnetic and paramagnetic contributions, 
and is thus very small \cite{byang}.  
B\"uttiker, Imry, and Landauer \cite{bu} argued that
persistent currents should, nonetheless, exist at the mesoscopic level,
and calculations with noninteracting electrons \cite{Ityp} predicted
typical currents of order $I_0\ell/L$ for a 
diffusive metallic ring, where $\ell \ll L$ is the elastic mean free path.  
Subsequently, persistent currents were observed experimentally in both
normal metal \cite{levy,webb} and semiconductor \cite{semi,semi2} rings.
Surprisingly, the persistent currents observed in metallic
rings were roughly two orders of magnitude larger than
those predicted by theories neglecting electron-electron 
interactions \cite{Ityp,Iave}, being of order $I_0$ in individual rings
\cite{webb}.  Mean-field calculations 
\cite{eckern,screening,screening2} for multichannel rings,
renormalization group results for 1D rings
\cite{1D-analytic}, and exact diagonalization studies \cite{2D-numerics}
found that the persistent current in disordered systems
can be enhanced by
repulsive interactions due to the suppression of charge fluctuations,
but were unable to account for the experimentally observed magnitude
of $I$.  

In this Letter, we propose a novel interaction effect which can lead to a large
enhancement of $I$ in multichannel rings, 
even in the ballistic regime, due to the fact that
the parities of different channels are no longer independent in an
interacting system. 
We investigate two integrable models of 
interacting $M$-channel rings, $SU(M)$ fermions with 
inverse-square $V(x)=g/x^{2}$ and 
delta-function $V(x)=U\delta(x)$ interactions.  In the prior model, 
we show that for $g >0$ the persistent currents
of all channels add constructively provided $k_F L > 2 \pi M$, 
leading to a large persistent current
whose parity depends only on the total number of electrons $N$.  In the 
$SU(M)$ delta-function gas, $I$ is found to oscillate in sign and increase in
magnitude with increasing $U$
due to a progressive condensation of electrons into the lowest subband,
leading to a parity-locking effect for
$U>16M^4 \hbar v_F/3\pi$, where $v_F$ is the Fermi velocity.
A disordered two-channel ring with interchannel
interactions is also investigated numerically, and shows,
importantly, that the parity-locking effect persists even when the subbands are
mixed strongly by disorder.  Qualitatively, the parity-locking effect 
arises in a thin ring because for sufficiently strong repulsive 
interactions, electrons can no longer pass each other.
The elimination of transverse nodes in the many-body
wavefunction (which are necessary for two electrons to pass each other) 
leads to a state whose parity depends only on the total number of electrons,
as discussed by Leggett \cite{leggett}, and whose persistent current is
consequently large.

Interacting spinless
electrons in a non-disordered ring with $M$ transverse channels,
threaded by an Aharanov-Bohm flux $(\hbar c/e)\phi$, may be represented by
1D $SU(M)$ fermions.  The transverse degrees
of freedom may be represented by an $SU(M)$ spin variable $\sigma=1,\ldots,M$. 
In the absence of
disorder, and for interactions which depend only on the electrons' 
coordinates along the ring ({\it i.e.}, for thin rings), the number of
electrons $K_{\sigma}$ in each channel is conserved.
The Hamiltonian of the system is 
\begin{equation}
H  =  -\frac{1}{2}\sum_{i=1}^{N} \frac{\partial^2}{\partial x_i^2} 
+ \sum_{i<j} V(x_i-x_j)
+ \sum_{\sigma=1}^{M} K_{\sigma} \varepsilon_{\sigma},
\label{ham.gen}
\end{equation}
where $N=\sum_{\sigma} K_{\sigma}$ is the total number of electrons and
$\varepsilon_{\sigma}$ is the energy minimum of subband $\sigma$.
Units with $\hbar=m=1$ are used.
The Aharanov-Bohm flux leads to the twisted boundary condition \cite{byang}
\begin{equation}
\Psi(x_1\sigma_1, \cdots, (x_i+L)\sigma_i, \cdots, x_{N}\sigma_{N})
= e^{i\phi}
 \Psi(x_1\sigma_1, \cdots, x_i\sigma_i, \cdots, x_{N}\sigma_{N}). 
\label{eq:boundary}
\end{equation}
For simplicity, let us consider equally spaced subbands $\varepsilon_{\sigma
+1} -\varepsilon_{\sigma}=\Delta\equiv E_F/M$.  The subband splitting $\Delta$
plays the role of an $SU(M)$ magnetic field.  
As we shall see, the effect of
repulsive interactions is to renormalize this effective field,
causing a condensation of electrons into the lowest subband.  
At $T=0$, the equilibrium persistent current is given by 
$I(\phi) = -(e/\hbar) \partial E_0/\partial \phi$, 
where $E_0(\phi)$ is the ground
state energy of Eq.\ (\ref{ham.gen}), subject to the boundary condition
(\ref{eq:boundary}).  For simplicity, we consider the case $\phi=\pi/2$
(1/4 flux quantum) in the following.

Let us first consider a model with long-range interactions:
$V(x)=g/d(x)^2$, where $d(x)=(L/\pi)|\sin(\pi x/L)|$ is the chord length
along the ring.  This model was introduced and solved by Sutherland
\cite{sutherland} for the case $M=1$ and $\phi=0$,
and is integrable for $g\geq 0$.
For $g=0$, this model reduces to
noninteracting electrons, and one finds
\begin{equation}
I(\phi_0/4) = \frac{e\hbar \pi}{2 m L^2}
\sum_{\sigma=1}^M (-1)^{K_{\sigma}} K_{\sigma}.
\label{I0}
\end{equation}
For $M\gg 1$, this leads to the well-known \cite{Ityp} result $|I| \sim
M^{1/2} I_0$ due to the random parities of the different channels.  
For $g>0$, on the other hand, the ground state is highly degenerate
in the limit $L\rightarrow \infty$ in the absence of $SU(M)$ symmetry
breaking ($\Delta= 0$) due to the strong repulsion of the potential
at the origin, which prohibits particle exchange.  In a finite ring, these
states differ in energy by at most $\pi \hbar v_F/L$ due to boundary effects;
all electrons will thus be condensed into the lowest subband for 
$\Delta>\pi\hbar v_F/L$, {\it i.e.}, for $k_F L > 2\pi M$, which is 
satisfied provided the ring is sufficiently thin.
The ground state of the system in this ``ferromagnetic'' state
has the Jastrow product form
\begin{equation}
\Psi (\{x\})
= \exp\left(i\frac{\phi-a}{L} \sum_{k=1}^N x_k\right) \prod_{1\le i< j\le N}
\left|\sin\left({x_i-x_j\over L}\pi\right)\right|^{\lambda} 
\sin\left({x_i-x_j\over L}\pi\right)
\end{equation}
for $0\leq \phi \leq \pi$, where
$a=0$ if $N$ is odd and $a=\pi$ if $N$ is even.
Here $\lambda=\sqrt{g+1/4}-1/2$.
One readily verifies that $\Psi$ is an eigenstate of Eq.\ (\ref{ham.gen}),
has the correct symmetry, and obeys the twisted boundary condition 
(\ref{eq:boundary}).  
This eigenstate is a positive vector when the particles are ordered, and
is therefore the ground state of the system.
The ground state energy is found to be
\begin{equation}
E_0(\phi)=\frac{\pi^2 (\lambda+1)^2 N(N^2-1)}{6L^2} + 
\frac{N}{2} \left(\frac{\phi-a}{L}\right)^2.
\end{equation}
The corresponding persistent current is 
\begin{equation}
I(\phi_0/4)= (-1)^N \frac{e\hbar \pi N}{2 m L^2} 
\sim (-1)^N M I_0.
\label{Ilock}
\end{equation}
The condensation of all electrons into the lowest subband caused by
the strong repulsive interactions
thus leads to an enhancement of the typical persistent
current by a factor of $M^{1/2}$ in the ballistic regime, due to the suppression
of the parity effect of Byers and Yang.

Let us next consider a model with short-ranged interparticle
interactions $V(x)=U\delta(x)$.  This system is integrable for arbitrary
$U$, and the eigenenergies with twisted boundary conditions may be determined
from a straightforward generalization of the nested Bethe ansatz of
Sutherland \cite{sun.delta} to the case $\phi\neq 0$.  
The energy of the system may be expressed as
\begin{equation}
E = \sum_{j=1}^N k_j^2/2 + \sum_{\sigma=1}^{M} K_{\sigma} \varepsilon_{\sigma},
\label{E.bethe}
\end{equation}
where the pseudomomenta $k_j$ are a set of $N$ distinct numbers which satisfy
the coupled equations
\begin{equation}
\exp[i(Lk_j-\phi)]=\prod_{\alpha=1}^{N-K_1}\frac{k_j-\Lambda^{(1)}_{\alpha}
+iU/2}{k_j-\Lambda^{(1)}_{\alpha}-iU/2}, 
\label{bethe1}
\end{equation}
\begin{equation}
\prod_{\beta=1}^{N_{n-1}}
\frac{\Lambda^{(n)}_{\alpha}-\Lambda^{(n-1)}_{\beta}-iU/2}{
\Lambda^{(n)}_{\alpha}-\Lambda^{(n-1)}_{\beta}+iU/2}
\prod_{\gamma=1}^{N_{n+1}}
\frac{\Lambda^{(n)}_{\alpha}-\Lambda^{(n+1)}_{\gamma}-iU/2}{
\Lambda^{(n)}_{\alpha}-\Lambda^{(n+1)}_{\gamma}+iU/2}
=-\prod_{\beta=1}^{N_{n}}
\frac{\Lambda^{(n)}_{\alpha}-\Lambda^{(n)}_{\beta}-iU}{
\Lambda^{(n)}_{\alpha}-\Lambda^{(n)}_{\beta}+iU},\,\,\,n=2,\ldots,M,
\label{bethe2}
\end{equation}
where $\Lambda_{\alpha}^{(n)},\,\alpha=1,\ldots,N_n=
N-\sum_{\sigma=1}^n K_{\sigma}$ are distinct numbers, with 
$\Lambda^{(1)}_j=k_j$.
For $\Delta=0$, the ground state is an $SU(M)$ singlet when $N$ is an odd
multiple of $M$.  As $\Delta$ increases, electrons in the higher subbands
are transferred to lower subbands, until all electrons are condensed into
the lowest subband for $\Delta > \Delta_c$.
This phenomenon is analogous to the spin-polarization
transition of the 1D Hubbard model in a magnetic
field, studied by Carmelo {\it et al}.\ and by Frahm and Korepin \cite{crit}.
Using the techniques of Ref.\ \cite{crit}, one finds 
\begin{equation}
\Delta_c = (1/4\pi)[U^2+(2\pi n)^2]\tan^{-1}(2\pi n/U) - Un/2,
\label{delta.crit}
\end{equation}
where $n=N/L$.  For $\Delta > \Delta_c$,
the system is in a parity-locked state, with persistent current given
by Eq.\ (\ref{Ilock}). 
It is useful to consider some limiting cases of Eq.\ (\ref{delta.crit}):
Using $k_F\simeq \pi n/M$, one finds 
$\Delta_c \simeq  M^2 E_F$ for $U = 0$ and $\Delta_c \simeq
16 M^3 E_F v_F/3\pi U$ for $U/v_F \gg M$.
For fixed $\Delta$, the critical interaction strength required
to enforce parity-locking is thus
\begin{equation}
 U_c \simeq 16 M^4 v_F/3\pi.
\label{Uc}
\end{equation}
For $M\gg 1$,
very strong interactions are thus required to cause complete parity-locking,
which would lead to a {\em macroscopic} persistent current.
This result is in contrast to that for the preceding model, which exhibited
parity-locking for any value of the interaction $g>0$, provided the ring was
sufficiently thin.

In order to see how the parity-locking effect develops as a function of
interaction strength, let us consider the simple case of a two-channel ring
with $N$ odd; then the parities of the two channels are necessarilly opposite.
For mesoscopic rings, $SU(2)$ excitations of the type considered by Kusmartsev
\cite{period1}
and by Yu and Fowler \cite{period2}, which lead to a 
$\phi_0/N$ periodicity of the persistent current, can be 
neglected \cite{period2}.  The persistent current for $M=2$ is then given
by
\begin{equation}
I(\phi_0/4) = (-1)^{K_1} (K_1-K_2)I_0/N.
\label{I2}
\end{equation}
For $\Delta \ll \Delta_c$, the polarization $(K_1-K_2)/L\simeq 2\chi\Delta$,
where the susceptibility $\chi$ may be evaluated
from Eqs.\ (\ref{E.bethe}--\ref{bethe2}) using the method of Shiba \cite{shiba};
one obtains $\chi= 2/\pi v_F$ for $U=0$ and
$\chi \simeq 3 U/(2\pi v_F)^2$ for $U\gg v_F$.
The magnitude of the persistent current is thus given by
$|I| \simeq 4e\Delta/\pi\hbar N$ for $U=0$, and is increased by the
factor $3U/8\pi v_F$ for $v_F \ll U \ll U_c$.  As $U$ is increased, the
persistent current thus oscillates in sign and grows roughly linearly
in magnitude due to the progressive transfer of electrons from the upper 
to the lower subband.
For $M\gg 1$, the evolution of the system toward the parity-locked state
as $U$ is increased
will of course be more complicated, but one nonetheless expects
 $I$ to fluctuate in sign and 
increase in magnitude as electrons condense into the lowest subband.

A peculiarity of the integrable models considered to this point is that the
number of electrons in each channel is a constant of the motion.  Both
disorder and more realistic interactions which depend on the transverse
coordinate will break this symmetry, and it is therefore important to 
verify that the parity-locking effect is not destroyed.  To this end, we
have considered a disordered two-channel ring, modeled in the tight-binding
approximation, with a nearest-neighbor interchain interaction $V$ included to 
induce interchannel correlations.
The Hamiltonian is
\begin{equation}
H= \sum_{i=1}^L \left[\sum_{\alpha=1}^2 
\left(e^{i\phi/L} c_{i\alpha}^{\dagger} c_{i+1 \alpha}
+ \mbox{H.c.}+\varepsilon_{i\alpha}\rho_{i\alpha}\right) + \frac{\Delta}{2}
\left(c_{i1}^{\dagger} c_{i2} + \mbox{H.c.}\right)
+V\rho_{i1}\rho_{i2}\right],
\label{ham.num}
\end{equation}
where $c_{i\alpha}^{\dagger}$ creates a spinless electron at site $i$ of chain
$\alpha$, $\rho_{i\alpha}\equiv c_{i\alpha}^{\dagger}c_{i\alpha}$, 
and $\varepsilon_{i\alpha}$ is a random
number in the interval $[-\varepsilon/2,\varepsilon/2]$.  
The interchain hopping determines the subband splitting $\Delta$.
Fig.\ 1 shows the persistent current for rings with 5 spinless
electrons on 18 sites  
as a function of $V$, calculated using the Lanczos technique.
Both the ensemble average of $I$ for 500 systems (squares) and the
persistent current of a typical system (solid curve) are indicated.
The error bars indicate the standard deviation $\delta I =
(\langle I^2\rangle-\langle I\rangle^2)^{1/2}$ over the ensemble (not the
statistical uncertainty in the mean value).
The subband splitting $\Delta=0.8$
is chosen so that in the absence of disorder and interactions, $K_1=3$ and
$K_2=2$, leading to a large cancellation of the persistent current due to
the different parities of the two channels.
The on-site disorder $\varepsilon=2>\Delta,\,E_F$ mixes the two
channels, but does not lead to strong backscattering.
For $V=0$, the sign of $I$ is essentially random, and $\langle I \rangle$
is slightly diamagnetic.  As $V$ is increased, $\langle I \rangle$
oscillates in sign and increases in magnitude, becoming strongly diamagnetic
for large $V$.  $|\langle I\rangle|$ is increased by a factor of 10 as
$V$ is increased from 0 to 20, while $\sqrt{\langle I^2\rangle}$ is 
increased by a factor of 3.
%
%
How can we
understand this behavior, so analogous to that of the ballistic system
with delta-function interactions described above,
given the strong intersubband scattering?  While
the subband occupancies are no longer constants of the motion,
there is a corresponding
topological invariant in the disordered system, namely,
the number of transverse nodes in the many-body wavefunction \cite{leggett}
({\it i.e.},
nodes which encircle the AB flux $\phi$).  The lowest subband has no
such nodes, while each electron in the second subband contributes one 
transverse node.  In order for two electrons to pass each other as they
circle the ring, such a transverse node must be present.  As $V$ increases,
it becomes energetically unfavorable for electrons to approach each
other, so transverse nodes in the many-body wavefunction will tend to
be suppressed.  In the strongly-correlated limit, all such nodes will be
eliminated, leading to a state whose parity depends only on the {\em total}
number of electrons. 
In such a state, the persistent currents
of all channels add constructively, leading to a large persistent
current (see Fig.\ 1).  It should be emphasized that while the enhancement
of the persistent current shown in Fig.\ 1 is relatively modest, a
much larger enhancement would be expected in a system with $M\gg 1$, based
on the above arguments.

In conclusion, we have proposed a novel interaction effect which leads to
a large enhancement of the persistent current in multichannel rings due
to correlations in the contributions of different channels.  It was shown that
even when interactions are weak compared to those necessary to enforce 
complete parity-locking, as is likely to be the case in
metallic rings such as those studied in Refs.\ \cite{levy,webb},
the persistent current may still be substantially enhanced by interchannel
correlations.  It should be emphasized that the parity-locking effect holds
in both ballistic and disordered systems; it is therefore
complementary to mechanisms previously proposed for the enhancement of the
persistent current \cite{eckern,screening,screening2,1D-analytic,2D-numerics}, 
which rely on the competition between disorder
and interactions.  It is likely that both such mechanisms are important to
explain the anomalously large observed value \cite{levy,webb}
of the persistent current in disordered metallic rings.

C.\ A.\ S.\ thanks M.\ B\"uttiker and F.\ Hekking for valuable discussions.
This work was supported by the Swiss National Science Foundation.

\begin{figure}
\vbox to 15cm {\vss\hbox to 17cm
  {\hss\
    {\includegraphics{plockfig2.eps}}
   \hss}
}
\caption{Persistent current $I=-(e/\hbar)\partial E_0/\partial 
\phi|_{\phi=\phi_0/4}$ of a disordered two-channel ring with 5 spinless
electrons on 18 sites as a function of the interchain interaction $V$. 
The current is given in units of $I_0=ev_F/L$.
The amplitude of the on-site disorder is $\varepsilon =2$ and the
subband splitting is
$\Delta=0.8$.  Solid curve: persistent current for one
realization of disorder.  Squares: ensemble average $\langle I\rangle$
for 500 systems.  The error bars indicate the width $\delta I=
(\langle I^2\rangle -\langle I\rangle^2)^{1/2}$ of the current distribution.
Note that the persistent current is diamagnetic for
large $V$, as expected for a system with $N$ odd due to parity-locking; 
this is true for all realizations of disorder.}
\label{system}
\end{figure}

\end{document}